# Methane present in an extrasolar planet atmosphere

Mark R. Swain*, Gautam Vasisht*, & Giovanna Tinetti†.

 *Jet Propulsion Laboratory, California Institute of Technology, 4800 Oak Grove Drive, Pasadena, California 91109-8099 USA. † Department of Physics and Astronomy, University College London, Gower Street, London WC1E 6BT, UK. ‡ Exoplanet and Stellar Astrophysics Laboratory, Code 667, NASA's Goddard Space Flight Centre, Greenbelt, Maryland 20771-0001, USA.

'These authors contributed equally to this work'

Molecules present in exoplanetary atmospheres are expected to strongly influence the atmospheric radiation balance, trace dynamical and chemical processes, and indicate the presence of disequilibrium effects. Since molecules have the potential to reveal the exoplanet atmospheric conditions and chemistry, searching for them is a high priority. The rotational-vibrational transition bands of water, carbon monoxide, and methane are anticipated to be the primary sources of non-continuum opacity in hot-Jovian planets[1,2,3]. Since these bands overlap in wavelength, and the corresponding signatures from them are weak, decisive identification requires precision infrared spectroscopy. Here we report on a near-infrared transmission spectrum of the planet HD 189733b showing the presence of methane. Additionally, a resolved water-vapour band at 1.9 μm confirms the recent claim[4] of water in this object. On thermochemical grounds, carbon-monoxide is expected to be abundant in the upper atmosphere of hot-Jovian exoplanets; thus the detection of methane rather than carbon-monoxide in such a hot planet[5,6] could signal the presence of a horizontal **chemical gradient away from the permanent dayside, or it may imply an ill-understood photochemical mechanisms that leads to an enhancement of methane.**

To date, molecular signatures have not been resolved in the emission spectra of hot-Jovian exoplanets[7,8,9,10]. Transmission spectroscopy during the primary eclipse (when the planet occults a portion of the stellar disk, and a fraction of light from the star is seen after traversal through the atmosphere around the planet's limb) has the advantage of being insensitive to temperature structure in the exoplanet's atmosphere[11]. Due to the presence of strong molecular absorption bands, near-infrared spectroscopy from 1–2.5 μm is well suited for detection of the signatures of $H_2O$, CO, and $CH_4$. For a hot-Jovian atmosphere in purely thermochemical equilibrium, the dominant carbon-bearing molecule is expected to be CO at higher temperatures ($T_{eq} > 1200$ K) and $CH_4$ at lower temperatures ($T_{eq} < 800$ K). On the daysides of the short-period, tidally-locked hot-Jupiters, the local carbon chemistry should be dominated by CO; disequilibrium effects may result in CO as the dominant carbon-carrying molecule, even on the terminator and nightsides of such planets[6]. Our detection of the onset of the $CH_4$ bandhead at 2.2 μm is the first clear spectral signature of a carbon-based molecule in an exoplanet atmosphere.

We observed the transiting exoplanet HD 189733b with the Hubble Space Telescope using the NICMOS (NIC-3) camera on May 25, 2007, over five contiguous spacecraft orbits. During these observations, NICMOS was configured to obtain a spectrophotometric time series between $1.4 - 2.5$ μm. As the parent star, HD 189733, is extremely bright (K = 5.52), we defocused the NICMOS camera to increase the saturation time of individual detector pixels. This provided additional benefits for precision photometry[12] by allowing some spatial averaging over intra-pixel and pixel-to-pixel quantum efficiency variations that could couple to rapid telescope pointing errors and slow pointing drifts to produce spurious intensity fluctuations in the measured time series. The defocus determines the spectral resolution allowing us to extract 18 independent wavelength channels with sufficient signal-to-noise. There are gaps in the measured time series because the star HD 189733 is not in the "continuous viewing zone" of the spacecraft, and, because of scheduling constraints, the in-eclipse data are not symmetric about the epoch of inferior conjunction.

During primary eclipse the apparent brightness of the star decreased by ~ 2.4 % (Fig. 1). However, as the modulation amplitude due to nominal amounts of water vapour[13] is predicted to be small (< 0.1%), high-dynamic-range (~$10^4$) spectra are required to detect and characterise any molecular features. Although the Hubble Space Telescope avoids the limitations imposed by the Earth's atmosphere, spectrophotometry with NICMOS or other on-board instruments[12,14] is subject to systematic errors that must be corrected because the errors are similar in size to the expected molecular signatures. In our data, the first orbit had strong systematic offsets (due to spacecraft settling) and was excluded from our analysis.

In order to arrive at the transmission spectrum it is important to establish a proper baseline for the remaining out-of-eclipse data. The systematics in the raw lightcurves are dominated by two types, correlations in time and in wavelength. In order to remove temporal correlations, we assumed that the observed flux in each wavelength channel for the out-of-eclipse orbits could be modelled by perturbations that were linear in five state variables, and by a term that was up to parabolic in spacecraft orbital phase. The state variables capture the optical state of the camera and are the centroids of channels, a variable defocus due to ``breathing'' of the telescope focus, rotation of the spectrum with respect to the detector, and temperature. A regression to the observed lightcurves provided the coefficients of the model. When decorrelated based on the model, the time series showed no further temporal correlations. Some remaining excess noise was strongly correlated in wavelength; an estimator for this noise was constructed as a weighted average of all channel time series data (collapsed in the wavelength axis) and was then subtracted from individual channels. The resulting channel time series are near detection noise limited. The robustness of the fits was verified by removing sections of the data from the fit and ensuring that the combined residuals remain well behaved.

The in-eclipse time series temporal effects were decorrelated by applying the model coefficients determined from the out-of-eclipse data to state variables and the spacecraft orbital phase at every time-step during that orbit. After accounting for the chromatic effects introduced in the eclipse light curve by limb darkening, a wavelength decorrelation was performed using the same method as was used for the out-of-eclipse orbits. The transit time series, now also corrected for instrument systematics, was averaged to construct a transmission spectrum. This spectrum includes astrophysical biases that have weak chromatic dependence, due to averaging over limb-darkened light curves and also due to the presence of cool starspots on the disk of the star. The effect of limb darkening, though not as dramatic as in optical data, is clearly seen as curvature in the eclipse light curves (Fig. 1). We co-added the spectral channels to construct synthetic light curves matching the H and K astronomical bands, for which there are published non-linear, limb darkening laws[15]. We then constructed a computer model of a limb-darkened star and simulated the motion of the planet across the stellar disk using the system's Keplerian parameters[14,16] to generate model light curves. A steepest descent scheme was used to fit generated curves to the measured data by leaving the planet effective radius (in that band) and epoch of transit as free parameters. The best fit light curves in each band were used to estimate the limb-darkening biases; the mean correction is $2 \times 10^{-4}$, and the colour correction is $1 \times 10^{-4}$ across the band.  The presence of starspots on HD 189733 has been inferred[16] from measurements
of long-term, out-of-eclipse variability, and from the structure of transit light curves in the
optical[14].  Unocculted starspots introduce a positive chromatic bias in the inferred
absorption depth, since they are cooler than the stellar disk; in the infrared, the chromaticity
of this effect is small and monotonic, and thus has little effect on the shape of the observed
modulation.  We derived a correction with a star spot coverage of 4 % using model
spectra[17] and assuming that the starspots were on average 1000 K cooler than the 5000 K
stellar photosphere.  A more detailed description of  the data reduction method is contained
in the Supplementary Information which is linked to the online version of the paper at
www.nature.com/nature.

We show the corrected spectrum as relative absorption depth in Figure 2.  The
signature of the $H_2O$ absorption band centred around 1.9 μm is immediately obvious.  The
steep increase in absorption at the short wavelength edge is also most probably due to an
adjacent water band centred shortward of 1.5 μm.  Thus, the new spectrum allows an
unambiguous identification of water vapour in the atmosphere of HD 189733b, confirming
its earlier inference from broadband photometry[4].  Since a steep change in absorption
occurs at 2.2 μm, the observations decisively show that methane is present in addition to
water.  To explore the abundance of $H_2O$ and $CH_4$ and possible contributions of CO and
$NH_3$, simulated transmission spectra were generated using a recent version of a planetary
spectral model[13].  The model covers a range of pressure from ~10 to ~$10^{-10}$ bar and includes
transitions[18] for $H_2$-$H_2$ (the most common molecular species).  The temperature and density
at each atmospheric level are determined by the P-T profile, and the absorption contribution
to each molecule is computed based on its mixing ratio. The $H_2O$ absorption coefficients
incorporate a new, high-accuracy list[19], and the CO absorption coefficients were estimated
with the HITEMP data base (L.S. Rothman et al., "HITEMP, the High-Temperature

Molecular Spectroscopic Database", private communication).  The $CH_4$ absorption coefficients were evaluated using a combination of line lists[20,21].  For all the molecules, the opacities are calculated for the selected spectral band at the different temperatures of the atmospheric layers (from 500 K to 2000 K) and in some cases[18,21] interpolated for intermediate values of the available temperatures.  We used the "evening terminator" pressure-temperature profile[22]; significantly different temperature profiles produce results that do not match the observed water vapour absorption features.  The theoretical spectra were binned to the same spectral resolution as the measurements, and the results of different compositions were compared with the observations using the reduced $\chi^2$ value.  Combinations of $H_2O$ and CO, as well as $H_2O$ and $NH_3$ failed to match the observed spectrum.  The model best fitting the observations has a mixing ratio of $\cong 5\times10^{-4}$ for $H_2O$, and $<5\times10^{-5}$ for $CH_4$ (see Fig. 2); the addition of $NH_3$ with a mixing ratio of $1\times10^{-5}$ improves the fit slightly.  The agreement between our $H_2O$ mixing ratio value and previous results[4] is significant because a wide range of wavelengths, covering three major $H_2O$ absorption bands, can be modelled self-consistently; this implies the estimated $H_2O$ mixing ratio is robust.  The pressure at which the atmosphere becomes optically thick ranges from a few millibars, when the absorption is strong, to ~0.2 bars, when the absorption is weaker (e.g. ~ 1.7 μm).  We have modelled the effect of aerosols and determined that our spectrum is haze-free, as their contribution would depress the spectral signature of both $H_2O$ and $CH_4$ relative to the measured absorption depth.  If aerosols are present, as is suggested by recent measurements[14], they must be in the form of small particles and their affects confined to wavelengths shorter than 1.5 μm[23].

While we can unambiguously determine $CH_4$ is present, the $CH_4$ abundance estimate is dependent on uncertainties with the high-temperature transitions.  Additionally, the presence of $CH_4$ masks the effect of CO in the absorption spectrum.  While the best fit is obtained using only $H_2O$ and $CH_4$, CO can be included up to the abundance of $H_2O$ with a modest increase in $\chi^2$. Thus, we cannot strongly constrain abundance of CO or the elemental C/O ratio. The relatively high concentration of $CH_4$ could be due in part to photochemistry. Photolysis of CO and $CH_4$ is probably not significant in the upper atmosphere (P < $10^{-4}$ bar) of HD 189733b because of the strength of C-O and C-H bonds and the deficit of ultraviolet flux from its cool parent star (spectral type K2V). However, the photolysis of the weaker bonds of sulphur, nitrogen, and oxygen-bearing compounds, and the accompanying availability of fast-reacting free radicals could have a significant impact on the relative abundances[24] of $CH_4$ versus CO.

**Acknowledgements** We thank Drake Deming for contributions to the original proposal and for suggestions for clarifying the material presented in the manuscript.  We thank Tommy Wiklind, Nor Pirzkal, and other members of the Space Telescope Science Institute staff for extensive assistance in planning the observations and for providing advice about ways in which the observations could be optimized.  We also thank Knud Jahnke for suggesting the long-wavelength NICMOS grism, Frederic Pont for discussions concerning the treatment of the data, M-C Liang for discussions on photolysis, Johathan Tennyson and Bob Barber for suggestions on the methane data lists.  G. Tinetti was supported by the UK Sciences & Technology Facilities Council and the European Space Agency.  The research described in this paper was carried out at the Jet Propulsion Laboratory, California Institute of Technology, under a contract with the National Aeronautics and Space Administration.

(Fig. 1) **Calibrated measurements showing the primary eclipse event.**  The measured time series (in units minutes of modified Julian day), corrected for systematic errors, of four orbits of the Hubble Space Telescope.  The primary eclipse occurs during the second orbit, and the curvature of the eclipse light curve is due to limb darkening of the star.  The two time series shown are co-added spectral data approximately matching the H and K astronomical bands (1.6–1.8 µm shown in blue & 2.0–2.4 µm shown in red, respectively).  The gaps in the time series are because HD 189733 is not in the continuous viewing zone for the Hubble Space Telescope.  For clarity, the red light curve has been offset.

(Fig. 2) **A comparison of observations with simulated water and methane absorption.**  The measured spectrum (black triangles), and two theoretical spectra of the predominantly $H_2$ atmosphere, showing the effects of small amounts of water (blue) and methane in combination with water (orange).  The measured spectrum contains significant differences at 1.7 – 1.8 µm and 2.15 – 2.4 µm from what is expected due to water vapour alone.  We interpret these departures as additional absorption features due to the presence of one or more molecules in addition to water.  When considering only water and methane, the theoretical spectrum best fitting the data was determined by binning the model (shown as white crosses) to the spectral resolution of the observations.  Different model predictions based on changing abundances and molecules were compared to the observations using the reduced $\chi^2$; the best fitting model has a water abundance of $5 \times 10^{-4}$ and a methane abundance of $5 \times 10^{-5}$.  The model spectrum can be improved slightly with the addition of small ($\sim 1 \times 10^{-5}$) amounts of either ammonia or carbon monoxide (shown in green and purple crosses, respectively).  The error bars for the measurements are plus and minus one sigma where the error includes the uncertainty in the correction of systematic effects (see Supplementary Information).  Note that determining the zero point for the spectrum depends on the diameter assumed for the planet and assumptions in the star spot correction. Thus, the shape of the spectrum is robust; there is an uncertainty of $\pm 2 \times 10^{-4}$ in the absolute level.

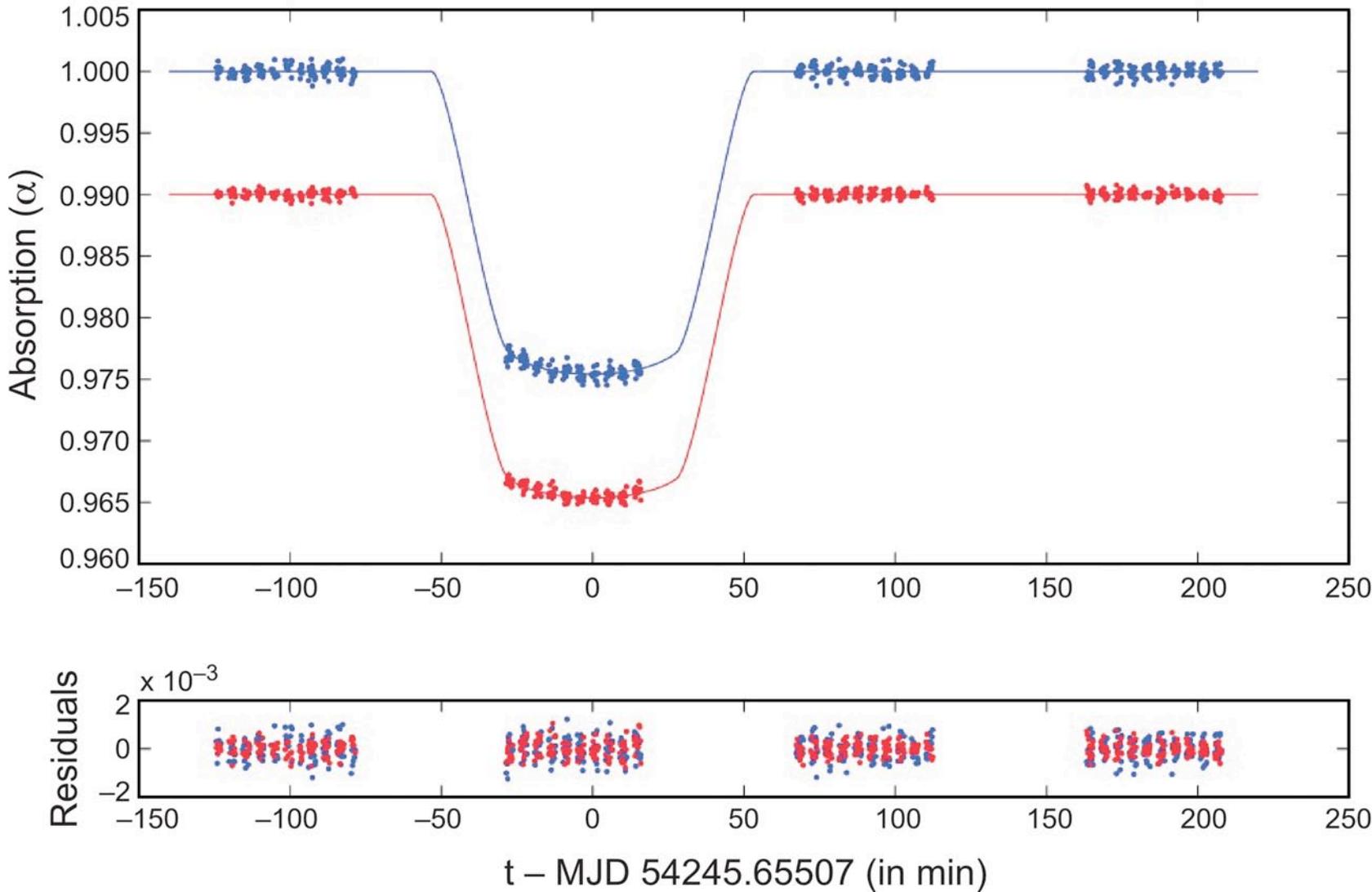

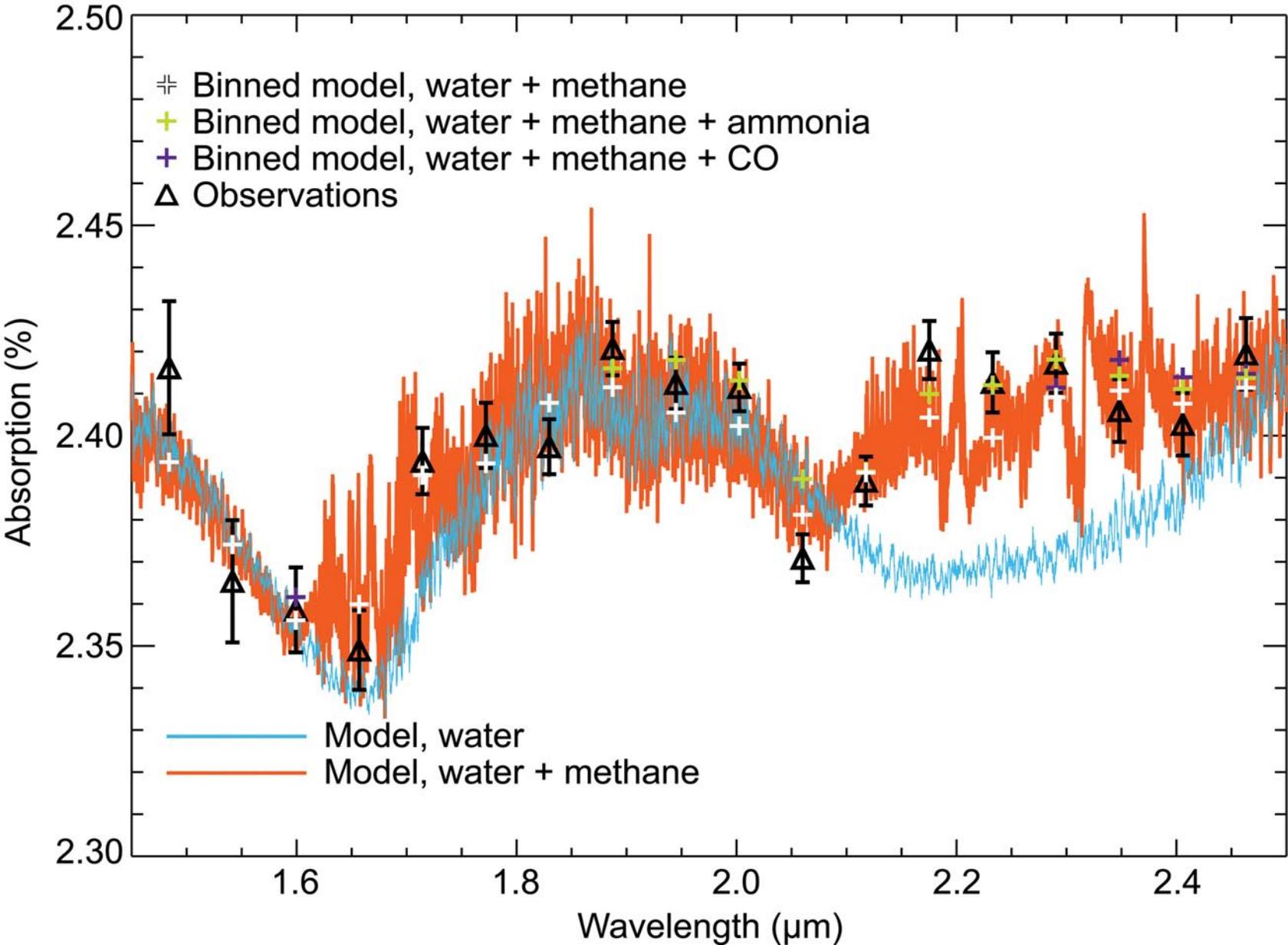